# Micromagnetic modelling of magnetic domain walls and domains in cylindrical nanowires

J.A. Fernandez-Roldan[1], Yu.P. Ivanov[2] and O. Chubykalo-Fesenko[1]

[1]Instituto de Ciencia de Materiales de Madrid, ICMM-CSIC. 28049 Madrid. Spain

[2]Department of Materials Science & Metallurgy, University of Cambridge, Cambridge CB3 0FS, United Kingdom

**Abstract**

*Magnetic cylindrical nanowires are very fascinating objects where the curved geometry allows many novel magnetic effects and a variety of non-trivial magnetic structures. Micromagnetic modelling plays an important role in revealing the magnetization distribution in magnetic nanowires, often not accessible by imaging methods with sufficient details. Here we review the magnetic properties of the shape anisotropy-dominated nanowires and the nanowires with competing shape and magnetocrystalline anisotropies, as revealed by micromagnetic modelling. We discuss the variety of magnetic walls and magnetic domains reported by micromagnetic simulations in cylindrical nanowires. The most known domain walls types are the transverse and vortex (Bloch point) domain walls and the transition between them is materials and nanowire diameter dependent. Importantly, the field or current-driven domain walls in cylindrical nanowires can achieve very high velocities. In recent simulations of nanowires with larger diameter the skyrmion tubes are also reported. In nanowires with large saturation magnetization the core of these tubes may form a helicoidal ("corkscrew") structure. The topology of the skyrmion tubes play an important role in the pinning mechanism, discussed here on the example of FeCo modulated nanowires. Other discussed examples include the influence of antinotches ("bamboo" nanowires) on the remanent magnetization configurations for hcp Co and FeCo nanowires and Co/Ni multisegmented nanowires.*

1. Introduction

Cylindrical magnetic nanowires constitute the most promising nanostructures for three dimensional nano-architectures in future applications for information technologies, nanoelectronics, sensing etc. From a more fundamental point of view they create a fascinating possibility for studying nanomagnetism in curved geometries which offers a plethora of novel static and dynamical properties [Fernandez-Pacheco2017, Stano2019]. Nanowires can be grown by chemical routes or more frequently by inexpensive electrodeposition in porous membranes with pore diameters ranging from several tens to hundreds of nanometers [Vazquez2015]. The lengths of nanowires are typically in the interval between 100 nm and 100 μm. This geometry presents a high aspect ratio length/diameter.

Among the properties of cylindrical nanowires which make them very appealing for future applications, we can summarize the most relevant:

- The elongated shape of cylinders induces a natural axial anisotropy via the magnetostatic energy effect which favors high stability of axial magnetic states. The curved geometry (via the exchange interaction) leads to an effective Dzyaloshinkii-Moriya interaction [Streubel2016] with no need to the intrinsic one.
- The overall effect of the intrinsic interactions and confinement in a curved geometry is to promote the formation of non-trivial topological structures in materials with no need of intrinsic Dzyaloshinskii–Moriya interaction or intrinsic anisotropies [Streubel2016]. Multiple novel magnetization textures have been reported for this geometry: toroidal, helicoidal, vortex (Bloch point) domain walls [Hertel2016a] and recently, skyrmion tubes [Charilaou2018, Fernandez-Roldan2018a].
- The dynamical vortex and skyrmion tubes may carry the Bloch point (a "hedgehog" skyrmion) [DaCol2014, Jamet2015, Hertel2016a, Charilaou2018]. The Bloch-point is a magnetic singularity with vanishing magnetization [Feldtkeller1965] which possesses some unique properties. The fascinating property of the Bloch point domain wall is that it is highly mobile [DaCol2014], with velocities higher than the transverse domain wall. Recently it has been shown that its motion leads to the emergent electromagnetic fields [Charilaou2018].
- Cylindrical nanowires are naturally magneto-chiral materials. For example, the magnetization mobility and stability of the vortex domain wall depend on its chirality [Hertel2013, Hertel2016a].
- Magnetic domain walls in cylindrical nanowires lack the so-called Walker breakdown, which consists of a sudden drop of the domain wall velocity under the action of applied magnetic field or current, observed in nanostrips [Hertel2013, Hertel2016a,]. However, the domain wall dynamics in cylindrical nanowires is limited by the spin-Cherenkov emission (spontaneous emission of spin waves by the domain wall for velocities higher than the minimum spin wave group velocity) [Hertel2016a]. This, however, happens at much higher velocities (up to 10 km/s) then the ones achieved in magnetic stripes. Therefore, ultrafast domain wall velocities can be expected.

Therefore, the cylindrical geometry offers multiple possibilities for observing highly non-trivial chiral magnetization structures in simple geometries. The main problem for nanowire applications is the control of domain wall dynamics: nucleation, mobility, pinning, etc. Only recently the development of the X-ray magnetic circular dichroism (XMCD) technique, highly sensitive magnetic force microscopy and electron holography has allowed to make first steps in observation of some of these structures such as circular magnetic domains [Ivanov2016b, Bran2016, Ruiz-Gomez2018] (extended vortex structure with a core parallel to the nanowire axis). Micromagnetic simulations play a decisive role in unveiling the observed magnetic contrast since they provide a complementary 3D information. The proper dynamical measurements on domain wall in cylindrical magnetic nanowires is a completely open field of research doing its steps [Ivanov2017, M. Schöbitz2019]. Most of studies report pinned domain walls, such as the

Bloch point domain wall [DaCol2014]. The strategies to pin domain walls involve the use of nanowires with different compositions [Kim2007, Ivanov2016, Garcia2018, Bran2018] or geometries [Dolocan2014, Arzuza2017, Chandra2014, Palmero2015, Mendez2018, Fernandez-Roldan2019] such as modulated nanowires or nanowires with notches or antinotches. Here micromagnetic simulations help to design nanostructures with desired properties and understand the pinning mechanism.

## 2. Hysteresis loops of magnetic nanowires

The hysteresis loops of magnetic nanostructures are strongly affected by their geometry through the shape anisotropy, and particularly, the aspect ratio length/diameter. Fig. 1 (a) illustrates the effect of the shape anisotropy in an ideal individual nanowire of Permalloy. Permalloy ($Fe_{20}Ni_{80}$) has a vanishing magnetocrystalline anisotropy, therefore, the behavior is dominated by the shape anisotropy effect. A preferential axial orientation of the magnetization resulting in a squared loop and a high remanence for the field applied along the nanowire axis is observed. The state before the switching (see Fig.2a below) is the two so-called open vortex structure, characterized by the predominant magnetization component along the nanowire direction but already developing a circular structure for the perpendicular components at the nanowire ends. The magnetization switching takes place in a unique Barkhausen jump. On the other hand, for a perpendicular magnetic field the hysteresis loop of a single nanowire exhibits a completely anhysteretic (rotational) behavior, i.e. the area enclosed between both branches of the hysteresis loop vanishes. Therefore, for a single nanowire of Permalloy the magnetic properties and the magnetization reversal process are fully determined by the shape anisotropy induced by the cylindrical geometry (magnetostatic energy). Ivanov et al. [Ivanov2013] showed that the magnetization reversal in other nanowires with fcc magnetocrystalline anisotropy in either polycrystalline or single crystalline nanowires is similar to Permalloy ones because it is also predominantly determined by the shape anisotropy owing to its large value compared to the magnetocrystalline anisotropy strength [Ivanov2013]. In the array of nanowire complex features may arise from the inter-wire dipolar interactions that decrease the effect of the shape anisotropy and introduce a "thin film" demagnetizing effect, see Ref. [Vivas2013, Fernandez-Roldan2018b, Stano2018].

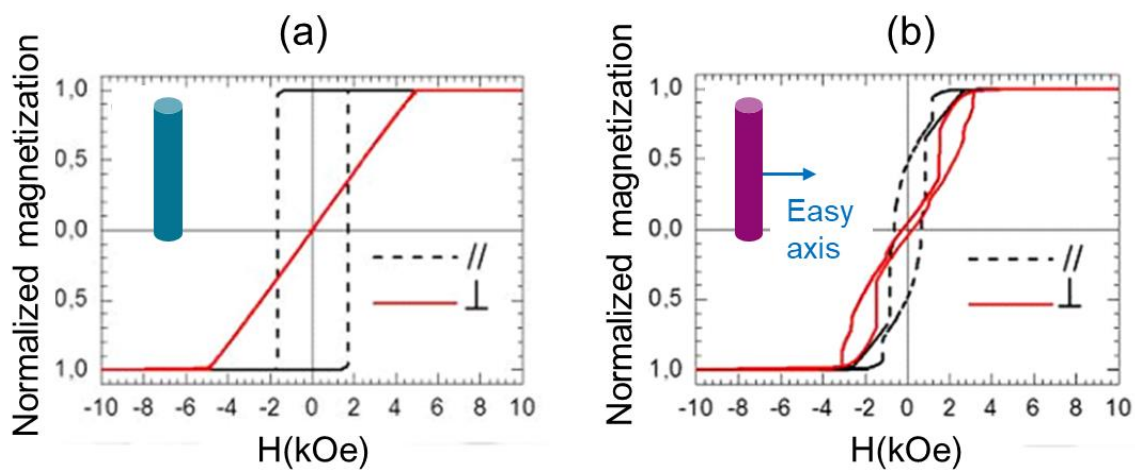

**Figure 1.** Simulated hysteresis loops for individual nanowires of (a) Permalloy with zero magnetocrystalline anisotropy (b) Co hcp with perpendicular magnetocrystalline anisotropy $4.5 \times 10^5$ J/m$^3$. Both nanowires have 40 nm diameter and 2 μm length. Figure adapted from [Ivanov2013]. Materials parameters are presented in Table I.

The introduction of a strong magnetocrystalline anisotropy, non-collinear with the nanowire axis, changes the hysteresis cycle due to the interplay between the involved energies. The uniaxial transverse anisotropy reported in Cobalt hcp nanowires exhibits a strong competition with the shape anisotropy [Ivanov2013, Ivanov2013b] almost completely suppressing it and producing a very soft behavior. For illustration, Fig. 1 (b) presents the simulated hysteresis loop of an individual Co nanowire with anisotropy easy axis perpendicular to the nanowire axis. Note also that the saturation magnetization of Cobalt is 75% higher than that of Permalloy, indicating a very high shape anisotropy. The inclined hysteresis loops for the single Co hcp nanowire in both parallel and perpendicular applied fields confirm that the magnetocrystalline anisotropy is strong enough to compensate the magnetostatic energy and to promote a drastic decrease of the coercive field and a small remanence. The remanent state is characterized by the formation of a vortex state along the whole length of the nanowire [Ivanov2013, Ivanov2013b], see Fig.2 (b) below. The perpendicular field hysteresis cycle resembles that of the vortex in soft magnetic dots characterized by the nucleation and annihilation of the vortices structure. The magnetostatic interactions between nanowires also significantly modify the hysteresis cycle due to the "thin film" shape anisotropy effect [Ivanov2013, Fernandez-Roldan2018b].

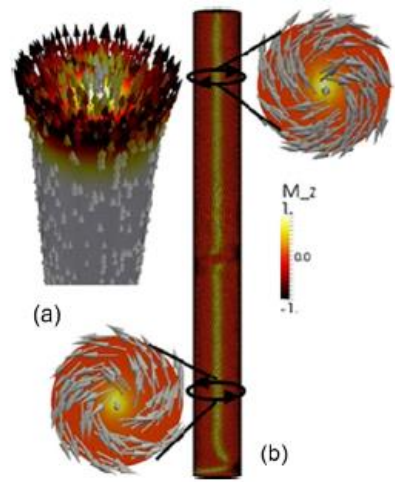

**Figure 2.** Magnetic configurations in cylindrical nanowire (a) precursor of the vortex domain wall (also known as open vortex structure) in shape anisotropy dominated nanowires. The image corresponds to the Permalloy nanowire at fields close to the switching one (b) the remanent state of hcp Co nanowires with easy axis perpendicular to the nanowire, showing two vortex domains with opposite chiralities. The field was applied parallel to the nanowire axis. Images taken from Ref. [Ivanov2013]. The same geometrical and magnetic parameters as in Fig.1 are used.

Experimental evidences and micromagnetic simulations confirm that the properties of the hysteresis loops of nanowires generally become independent of the nanowire length for aspect ratio values larger than 8 – 10 [Bance2014, Vivas2011]. As an illustration,

Figure 3 displays the calculated hysteresis loops of Permalloy and Co fcc nanowires with different lengths. For short nanowires, a lower coercive field is observed and the hysteresis loops are strongly affected by small variations of the length. The anisotropy of the fcc phase of Co has also a remarkable effect in the inclined hysteresis loops for short nanowires. In fact, 30 nm length Co nanowire is a pillar which presents the characteristic vortex hysteresis cycle. On the contrary, for nanowires with large aspect ratios, the coercive field and the remanence become practically independent on the length. Note that in this ideal case in nanowires with lengths smaller than ca. 70 nm (estimated for fcc Co) the open vortices with the same chiralities are formed at the opposite ends due to the minimization of the magnetostatic interaction between them. When the length is increasing, the interaction of structures at opposite ends becomes negligible and the vortices at the opposite ends are formed with the opposite chiralities as in Fig.2b. The latter happens due to the influence of the nanowire internal demagnetizing field (with an inward and outward component at opposite nanowire ends), resulting in opposite magnetic torques. The latter behavior is common in simulations for all nanowires in the absence of defects and granular structure.

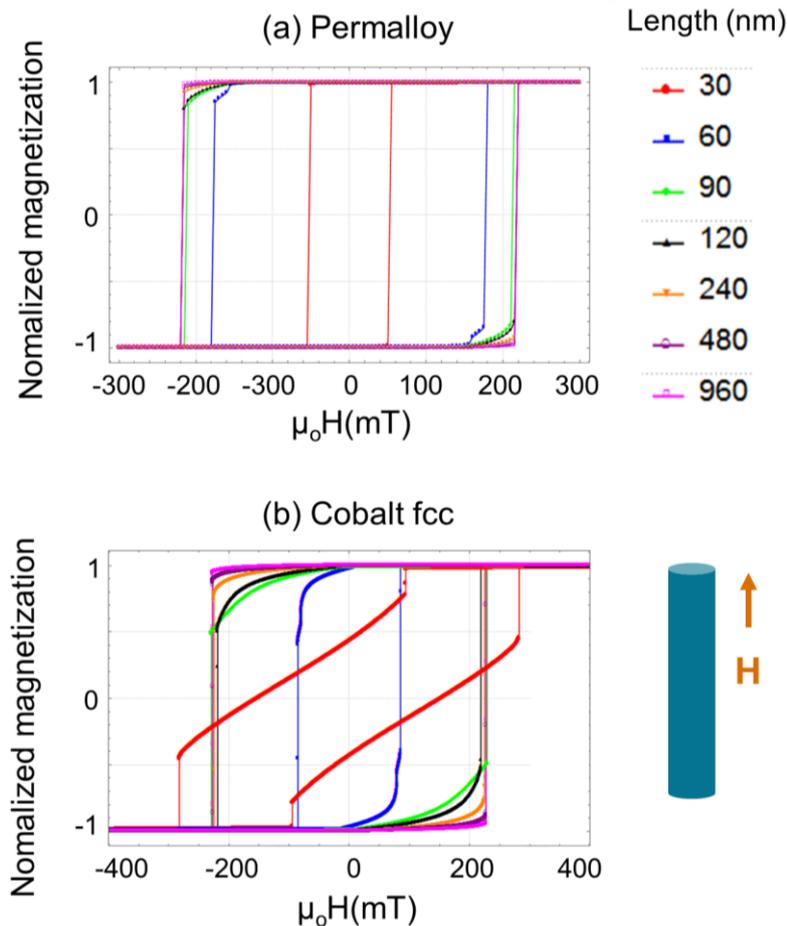

**Figure 3.** Simulated hysteresis loops for individual nanowires with several lengths without and with magnetocrystalline anisotropy. Figures correspond to individual nanowires of (a) permalloy and (b) Co fcc with a 30 nm diameter. Parameters used for simulations can be found in Table I.

The nanowire diameter has a more critical impact on its hysteresis properties than the length. As it is presented in Fig.4, the coercive field strongly decreases with the increase of the NW diameter which reflects the change of the reversal mode from that of the transverse domain wall to the Bloch point (vortex) domain wall [Ivanov2013], see next section. At the same time, the remanence has a small dependence on the diameter, defined by the formation of the open vortex structures for shape anisotropy-dominated nanowires, see Fig.2a. This dependence becomes more pronounced for nanowires with high saturation magnetization value such as Fe-based nanowires (for example Fe bcc or FeCo nanowires) [Ivanov2013].

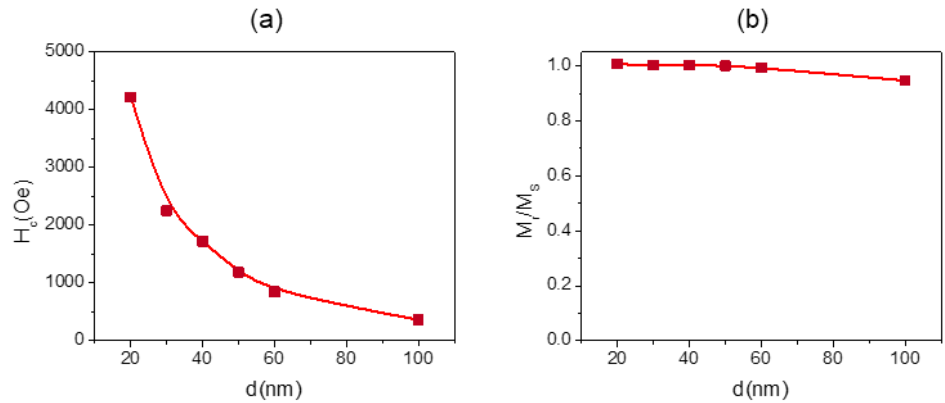

**Figure 4.** (a) Coercive field and (b) remanence dependence on the nanowire diameters for individual Permalloy nanowires.

## 3. Magnetic domains and domain walls in straight magnetic nanowires

As it is indicated in the previous section, shape anisotropy-dominated magnetic nanowires are in the (almost) single domain state in the remanence and demagnetize via the domain wall propagation [Forster2002a, Hertel2002, Forster2002b, Ivanov2013]. The type of domain wall by which the nanowire demagnetizes is determined by the nanowire geometry and material [Ivanov2013]. More concretely, only nanowires with very small diameters demagnetize via the transverse domain wall propagation (see Fig.5a). Depending on the relative orientation between domains, the wall is called tail-to-tail or head-to-head transverse domain wall [Jamet2015]. The transverse domain wall carries a magnetic charge in volume due to the divergence of magnetization [Jamet2015].

Larger-diameter (or saturation magnetization) nanowires demagnetize via the Bloch point (vortex) domain wall propagation, see Fig.5b. The vortex domain wall presents an azimuthal magnetization component with the flux-closure structure. In vortex structures, the relative orientation between its azimuthal component and the core direction defines the chirality of the vortex. The direction of the core indicates the polarity. In the case of Co hcp nanowires with perpendicular diameter, for diameters equal and larger than 60 nm the nucleation of vortex structures along the whole nanowire length takes place. The vortex domain wall typically carries a mathematical singularity called Bloch point [Forster2002a, Hertel2004, Kim2013 ], which is s unique topological defect in ferromagnetic materials [Wartelle2019], where the magnetization vanishes as displayed in Figure 5(b). Therefore, it eventually received the name of Bloch point wall to differentiate it from the (surface) vortex wall observed in ultrathin nanostrips [Jamet2015]. This singularity has been already experimentally observed in nanowires by XMCD-PEEM technique [DaCol2014]. The transition diameter between the two reversal modes decreases as the nanowire saturation magnetization value increases, namely for permalloy this happens for diameters ca. 40 nm, for Co fcc nanowires for diameters ca. 20nm, and the Fe-based nanowires generally demagnetize via the vortex domain wall propagation [Ivanov2013]. The Bloch point domain wall in reality carries a 3D hedgehog skyrmion, see Fig.5b. Note that the vortex core is not necessary located in the nanowire center, but may form a spiral [Fernandez-Roldan2018a]. Finally, the proper Neel-type hedgehog 3D skyrmion structure can be also formed as a stable configuration in small diameter nanowires with very large saturation magnetization such as FeCo, see Fig. 5c.

There is another name for a frequently metastable domain wall for nanowire diameters close to the transition value between the vortex and the transverse domain walls, depicted in Figure 1.14(e). This domain wall type presents characteristics of both previous wall types and is called asymmetric transverse domain wall [Ferguson2015, Jamet2015], see Fig.5e. In reality it corresponds to the situation when the vortex core goes out from the nanowire center to its surface [S. Komineas1996]. Together with the vortex on the surface of nanowire, there is an anti-vortex structure opposite to it to preserve the topological charge. Recently, the topological transformation from asymmetric transverse domain wall to a vortex domain wall in cylindrical nanowires, mediated by Bloch point injection, has been reported [Wartelle2019]. Other topological transformation between magnetization textures are also theoretically possible by Bloch point ejection [Wartelle2019].

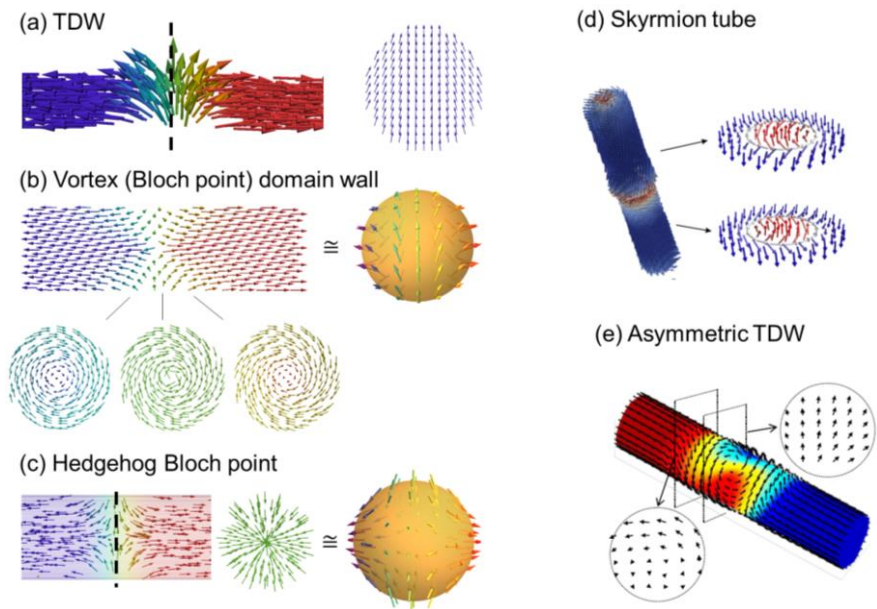

**Figure 5.** Domain walls in nanowires (Permalloy). (a) Transverse domain wall (TDW). (b) Vortex-Bloch point domain wall. (c) Hedgehog Bloch point separating two magnetic domains. (d) Skyrmion tube from Ref. [Charilaou2018] formed during the magnetization reversal. (e) Asymmetric transverse (vortex) domain wall from Ref. [Ferguson2015]. Cross-sections indicate the inner magnetization of the domain walls at the marked positions.

For nanowires with larger diameters ca. 100 nm another reversal mode called recently a skyrmion tube is possible, especially when the saturation magnetization is high. Differently to the nanowires with smaller diameter it does not carry a Bloch point but consists of a core-shell structure. The outer shell first rotates towards the field direction while the nanowire core remains magnetized in the opposite direction, see Fig.5d. The skyrmion tubes were noticed [Charilaou2018] in Permalloy during the dynamical demagnetization process. The magnetization in nanowire cross-section indeed has the same topological structure as the classical Bloch skyrmions but here induced by the curvature of the nanowire in the absence of the Dzyaloshinskii-Moriya interaction [Dzyaloshinsky1958, Moriya1960]). During further demagnetization, the inner skyrmion tube core breaks and a hedgehog skyrmion (Bloch point) is formed. Its propagation completes the demagnetization process. We should stress that these structures in Permalloy are dynamical. In FeCo nanowires (high saturation magnetization) they are shown to be stable at negative applied fields [Fernandez-Roldan2018a].

Most of domain walls are dynamical objects, however, they can separate different domains and can be pinned at the defects, thus being stable structures. As many promising applications rely on the nucleation and control of the magnetization processes and magnetic domain walls, there is a strong effort in the nanomagnetism community to understand and control their pinning mechanisms [Kim2007, Ivanov2016, Garcia2018, Bran2018, Dolocan2014, Arzuza2017, Chandra2014, Palmero2015, Mendez2018, Fernandez-Roldan2019].

Conventionally, a magnetic domain has been considered as a uniformly magnetized region in the mesoscale in thermodynamic equilibrium. Then, a magnetic state was defined as a collection of magnetic domains. However, the 3D nature of nanowires and the emerging properties at the nanoscale, make us to redefine the concept of magnetic domains as stable configurations of the magnetization which may lack of homogeneity due to the nanowire curvature. For example, we define the part of the nanowire in the vortex state as a vortex domain. Therefore, some materials with large saturation magnetization and high transverse magnetocrystalline anisotropy (as Co hcp nanowires) may show a vortex configuration along the nanowire length as the ground state [Ivanov2013b, Ivanov2013]). In multisegmented nanowires, depending on the segment size and material, a single vortex, two vortex or multivortex domain states (see Fig. 6 below) have been reported [Ivanov2013b, Bran2016, Ruiz-Gomez2018]. For example, Fig. 2b shows a two-vortex domain, with two opposite vortex chiralities. Depending on the micromagnetic parameters and crystalline structure of the nanowire either vortex domains, see Fig.7 or transverse domains, see Fig. 7a or their combination is possible, see Fig.7b. The latter combination has been observed in $Co_{85}Ni_{15}$ nanowires by XMCD and confirmed by micromagnetic simulations [Bran2017]. This shows that the domain structure is not unique and for some parameters set transverse and vortex domains are metastable states. Another possible domain configuration discussed in Ref. [Lebecki2010] is alternating vortex structures with core on the nanowire surfaces formed between the transverse domains.

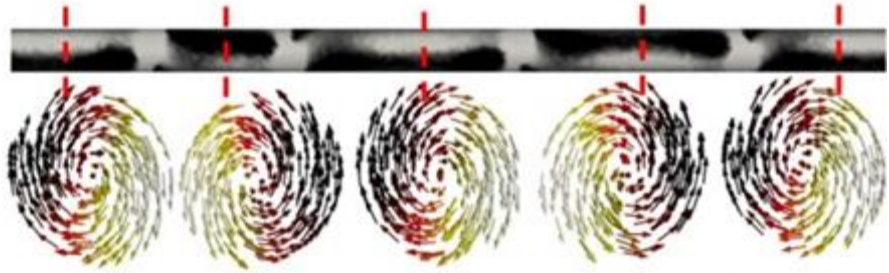

**Figure 6**. A multivortex domain state with opposite vortex chiralities along the nanowire. The calculated image corresponds to Cobalt nanowire with perpendicular anisotropy. Figure extracted from Ref. [Ivanov2016b]. Paramenters used in simulations are listed in Table I.

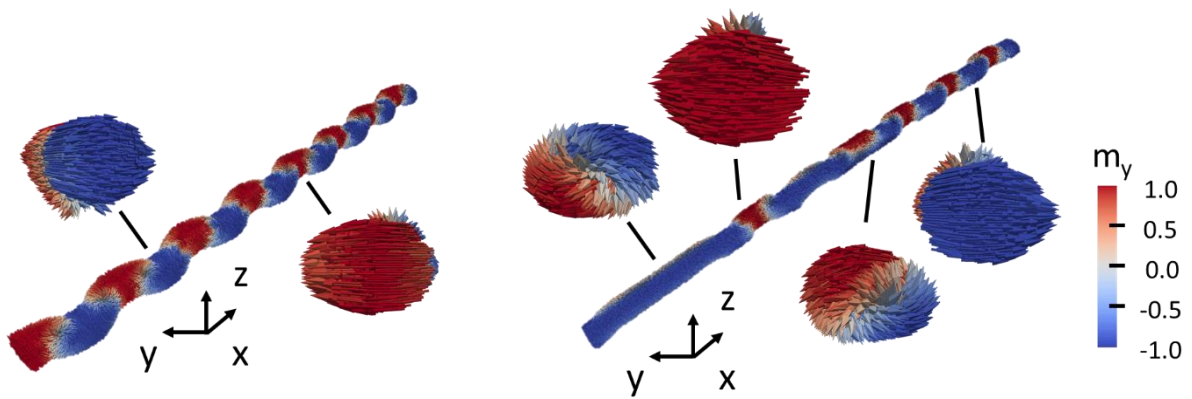

**Figure 7.** Calculated magnetic domain pattern in $Co_{85}Ni_{15}$ nanowires (a) transverse domains (b) alternating transverse and vortex domains. Micromagnetic parameters used in modelling are found in Table I.

### 4. Domain wall velocity in cylindrical magnetic nanowires

One of the most fascinating characteristics of cylindrical nanowires are the dynamical properties induced by their cylindrical symmetry and magneto-chiral effects already mentioned in previous sections. The propagation of the domain walls driven by applied magnetic fields and currents in nanostrips is limited in velocities of the order of 100 m/s (in the absence of the Dzyaloshinkii-Moriya interactions). The steady velocity achieved with the driving force is initially increasing (with the increase of the field or current) and then drops at the so-called Walker breakdown field. This phenomenon limits the speed of future information technologies. Theoretical considerations demonstrate that in cylindrical nanowires the Walker breakdown [Schryer1974] does not exist due to the cylindrical symmetry [Thiaville2006, Yan2011] or at least it is displaced to higher velocity values. Fig. 8 presents the calculated domain wall velocity in a cylindrical Permalloy nanowire as a function of the applied field showing velocities almost up 10 km/s, saturating at high field values. This saturation occurs due to the so-called spin-Cherenkov effect [Yan2011, Hertel2016a] when the domain wall loses energy with the spin-wave emission.

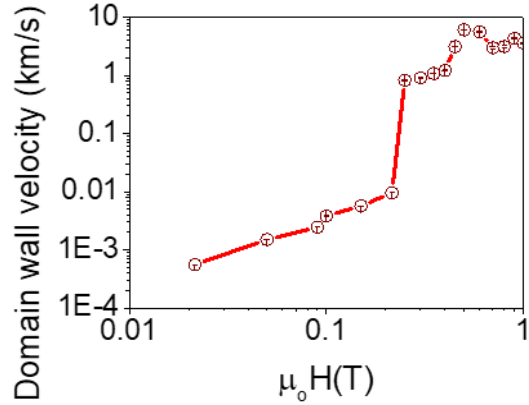

**Figure 8.** Calculated velocity of a vortex (Bloch point) domain wall in a Co fcc nanowire with diameter of 40 nm showing saturation due to the spin wave emission. Micromagnetic parameters used in modelling are summarized in Table I.

As for the influence of the current, calculations indicate that the direct motion of the vortex domain wall with current requires too high current and the assistance with applied field will be probably necessary. The Oersted field being circular in nature can change of the chirality of the vortex domain wall [M. Schöbitz2019].

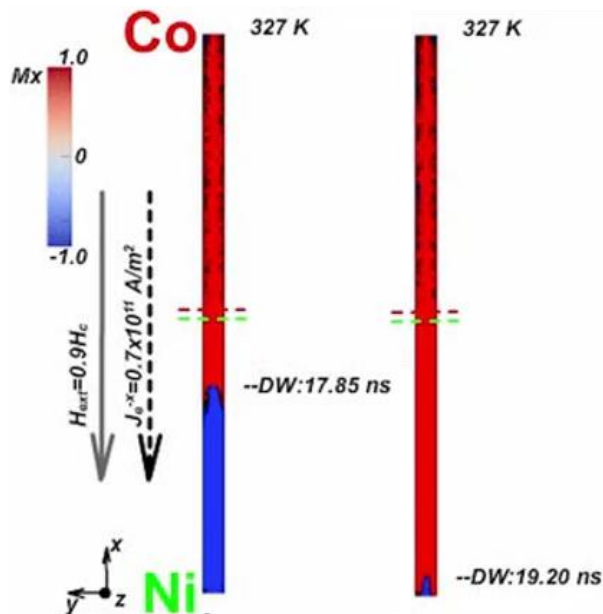

**Figure 9.** Micromagnetic model for domain wall propagation for segmented Co/Ni nanowire under applied magnetic field and current at two different times. Dashed lines indicate the interphase between Co and Ni segments. Modelling parameters are summarized in Table I.

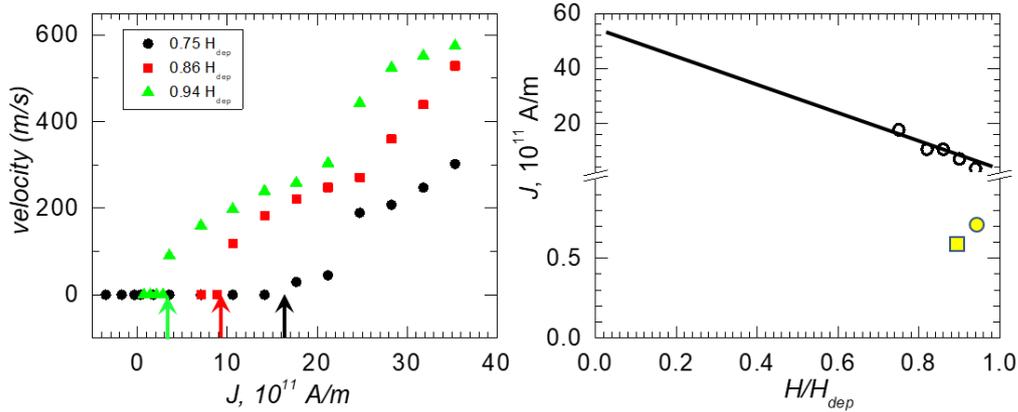

**Figure 10.** (a) Domain wall velocity under different current densities J for various external magnetic fields values with respect to the de-pinning field $H_{dep}$ simulated at 0 K temperature (arrows show the critical current density for DW depinning). b Dependence of the critical current density on the external magnetic field values at 0 K (open circles), at 327 K (filled circle) and experimental data at 327 K (filled square). The line shows linear approximation towards zero field DW motion. Images from Ref. [Ivanov2017].

In the Ref. [Ivanov2017] the current-driven VDW dynamics in cylindrical nanowires were studied by in-situ Lorentz TEM in the presence of the external magnetic field to promote the DW dynamics for the lower value of the applied current densities. The cylindrical nanowire of 80 nm diameter, 20 μm length and consisting of segments of Co and Ni of 700 nm each was used. Additionally, the experiments have been performed only for the case of the constant current which is ruled out the possibility to measure the DW velocity. The interfaces between segments of different magnetic materials was used as an effective site for the domain wall pinning (more details in the section "Domain walls in multisegmented Co/Ni nanowires"). Consequently, the value of the external magnetic field needed to nucleate DW in nanowire, the field needed to de-pin the DW from the interface between the segments as well as the critical current density for the DW propagation were determined. Those parameters were compared with the ones obtained by the micromagnetic simulations of a 80nm diameter nanowire with 2 segments of 700 nm each, see model in Figure 9. The direction of the magnetocrystalline anisotropy was considered in agreement with TEM data (at 850 with respect to the NW axis for the Co segment and {220} texture for Ni segments. The spin polarization of 0.42 reported for planar structures [Boulle2011] was used. The simulations show that these nanowires demagnetize via the vortex domain wall propagation. The simulations has been done for 0 K and for finite temperatures (extracted from the experiments).

Fig. 10 presents a simulated velocity data for the domain wall propagation from Co to Ni segment as a function of the current density J for several applied field values. It shows the existence of a critical value of $J_c$ above which current-driven DW motion occurs, and this value depends on the applied external magnetic field. The maximum value for the domain wall speed was found to be as high as 600 m/s in agreement with theoretical predictions. Simulations show that the critical density of the spin-polarized current at 0 K decreases almost linearly with the increasing of the external magnetic field (Fig. 10b). Note that the simulated value of $J_c$ at 0 K is more than one order of magnitude larger than the experimental data. The $J_c$ value calculated for 327 K (the final state is shown as a triangle in Fig. 9b) is very close to the one obtained experimentally (red cross in Fig.

10b). Most probably there are additional contributions to the current driven 3D DW motion which were not taken into account in the micromagnetic model. The complete understanding requires more experiments with nanoseconds DW dynamics and advanced ultra-fast imaging technique.

## 5. Domains and domain walls in nanowires with geometrical modulations

Domain-wall-based technologies as the racetrack memory [Parkin2015] require a precise control over the magnetization and domain wall location. In order to manage the positioning of domain walls along the nanowire length, multiple strategies have been proposed to achieve the local pinning of the domain wall. The fundamental idea is the creation of potential wells and barriers where the domain wall may eventually get pinned. Geometrical strategies consist of creating constrictions along the nanowire length, artificial notches, antinotches, defects or diameter modulations which act as pinning sites for the domain wall [Dolocan2014, Arzuza2017, Chandra2014, Palmero2015, Mendez2018, Lage2016]. In addition, variations of the material along the nanowire length or at specific locations [Kim2007, Ivanov2016, Garcia2018, Bran2018] also offer the possibility of promoting domain wall pinning. Micromagnetic simulations play here a decisive role since they allow to rapidly vary geometrical and material parameters with the aim to design and understand pinning. Here we discuss two possibilities of geometrical constrictions based on modulations in diameter (with alternating diameters), see Fig.11a, and geometrical notches, see Fig.11b in cobalt-based nanowires. The micromagnetic parameters correspond to Co and FeCo, see Table I.

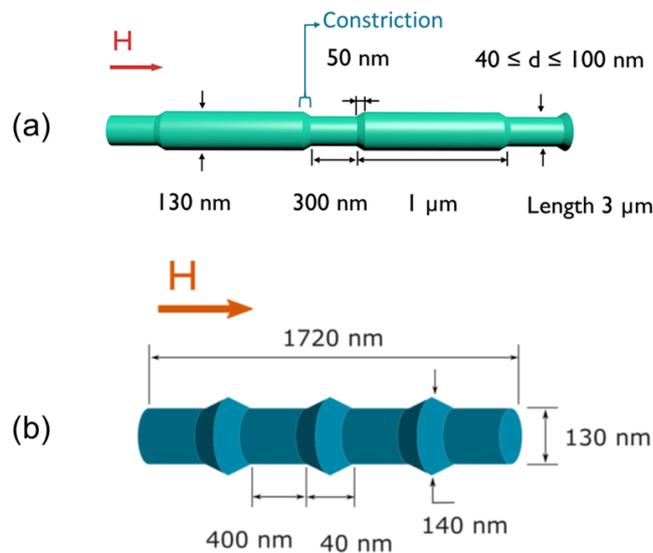

**Figure 11.** Geometry of simulated (a) nanowire with geometrical modulations and (b) "bamboo" nanowires.

| Material | $\mu_o M_s$(T) | $A_{ex}$(pJ/m) | $l_{ex}$(nm) | Crystal structure | Anisotropy type | Easy axis | $K_1$(kJm$^{-3}$) | $K_{sh}$(kJm$^{-3}$) |
|---|---|---|---|---|---|---|---|---|
| Permalloy (Fe$_{20}$Ni$_{80}$) [Ivanov2013] | 1.0 | 10.8 | 5.2 | Polycrystalline | - | - | 0 | 199 |
| Co(111) [Ivanov2013] | 1.76 | 13.0 | 3.3 | FCC | Cubic | [1,1,1] \|\| nanowire axis | -75 | 616 |
| Co(100) [Ivanov2013] | 1.76 | 13.0 | 3.3 | HCP | Uniaxial | At 75º with nanowire axis | 450 | 616 |
| Co-hcp [Ivanov2013, Moreno2016] | 1.76 | 30.0 | 4.9 | HCP | Uniaxial | At 75º with nanowire axis | 450 | 616 |
| Fe$_{30}$Co$_{70}$ [Bran2013] | 2.0 | 10.7 | 2.6 | BCC Polycrystalline | Cubic | A random in-plane easy axis component considered in each grain | 10 | 796 |
| Ni(111) [Ivanov2013] | 0.61 | 3.4 | 4.8 | FCC | Cubic | [111] \|\| nanowire axis | -4.8 | 74 |
| Co$_{85}$Ni$_{15}$ [Bran2017, Kronmuller1997, Moreno2016, Vega2012, Moskaltsova2015] | 1.60 | 26.0 | 5.1 | HCP | Uniaxial | At 88º with nanowire axis | 350 | 509 |
| Co$_{65}$Ni$_{35}$ [Bran2017, Kronmuller1997, Moreno2016, Vega2012, Moskaltsova2015] | 1.35 | 15.0 | 4.5 | HCP Polycrystalline | Uniaxial | At 65º with nanowire axis | 260 | 362 |
| Co$_{35}$Ni$_{65}$ [Bran2017, Kronmuller1997, Moreno2016] | 1.01 | 10.0 | 5.0 | FCC | Cubic | [111] \|\| nanowire axis | 2 | 203 |

**Table I**. Materials parameters used in micromagnetic modelling, microstructure and derived quantities: Saturation magnetization $\mu_o M_s$, exchange stiffness $A_{ex}$, exchange length $l_{ex}=(2 A_{ex} / \mu_o M_s^2)^{1/2}$, crystal structure, magnetocrystalline anisotropy type, easy axis direction, first magnetocrystalline constant $K_1$ and shape anisotropy $K_{sh}= \mu_o M_s^2/4$. Santuration magnetization and exchange stiffness from CoNi alloys have been obtained by linear interpolation with the Co content of the alloy in Ref [Bran2017].

The modulated nanowires were considered made of FeCo material with high saturation magnetization value. To represent realistic experimental situations, the granular structure was modelled by Voronoi construction and several disorder distributions were considered. Hysteresis loops for modulated nanowires under a magnetic field, applied parallel to the symmetry axis of the nanowire depend considerably on the particular polycrystalline nanostructure of the nanowire. As a main characteristic, the hysteresis loops become less squared, the remanence increases and the switching field decreases as the minor diameter of the nanowire becomes larger, see Fig.12. The demagnetization process always starts in the segments of the larger diameter by means of the nucleation of the open vortex structures at the constriction. These structures depin from the constriction and propagate first inside the wide diameter segments and later inside the small diameter segment. When the difference between diameters is small the propagation in both segments, although consequent and not simultaneous, takes place at the same field value ('weak' pinning). In the opposite case of large diameter difference ("strong pinning"), the hysteresis loop consists of a propagation stage inside the large dimeter segment and depinning (switching) of the magnetization from the constriction to propagate inside a small diameter segment occurring at a different field.

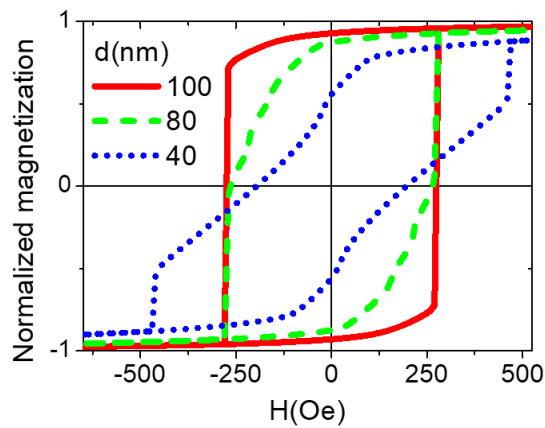

**Figure 12.** Example of simulated hysteresis cycles for individual modulated FeCo nanowires with three different minor diameters for some particular disordered distribution. Modelling parameters are collected in Table I.

Importantly the open vortex structures (similar to presented in Fig2a) at the constrictions between large and small segments are formed with arbitrary chirality. Different realizations of granular structures promote different chirality patterns, therefore in one particular segment vortices at the opposite ends may be formed with the same or different chirality. These vortices propagate towards the center of the large diameter and span almost the whole segment. In the case of opposite chirality, the propagation leads to a topologically protected structure which requires a large field to annihilate. As the field progresses, the spins in the outer shell rotate towards its direction forming a skyrmion tube, see Fig.13 b. Indeed, the magnetization structure in the nanowire cross section is formed by a magnetic skyrmion with the core pointing against magnetic field direction and the shell pointing parallel to it.

In order to prove that the structure is indeed a skyrmion, we evaluate in Fig.15 the topological charge as a surface integral evaluated in the nanowire cross-section along its length

$$Q = -\frac{1}{4\pi} \int_S \vec{m} \cdot \left(\frac{\partial \vec{m}}{\partial x} \times \frac{\partial \vec{m}}{\partial y}\right) dS$$

Fig.14 presents the topological charge along the nanowire for two disorder realization and large diameter difference. In the case of Fig.14a the vortices at the ends of the large diameter segments are formed with the same chirality, as the field is progressing they are transformed into the skyrmion tubes with topological charge almost one in the center of the nanowire. In the case of Fig.14 b the vortices in the left segment were formed with the opposite chirality, when the field is increasing in the opposite direction the two tubes meet and the topological charge in the middle of this segment is almost zero, reflecting the topological protection. This situation needs additional field in order to annihilate and complete the magnetization reversal.

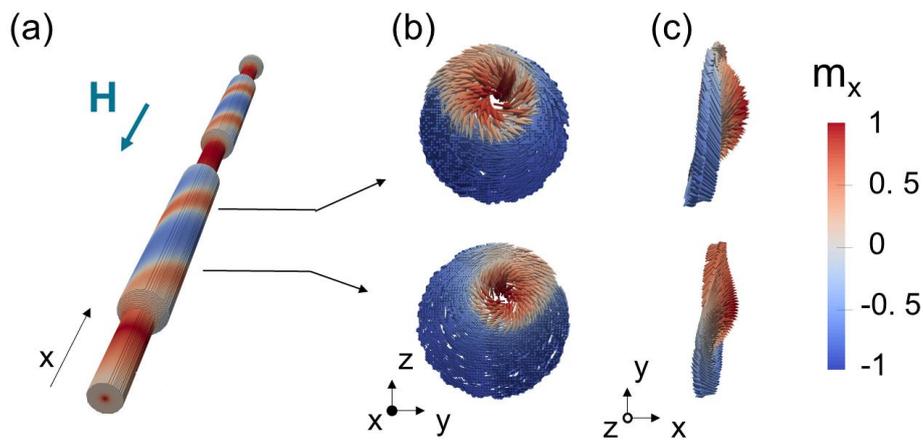

**Figure 13.** (a) The surface magnetization configuration of the multisegmented FeCo nanowire showing the helicoidal ("cork-screw") structure (b) The magnetization distribution in the cross sections of the larger segment at indicated positions, showing the skyrmions with displaced core.

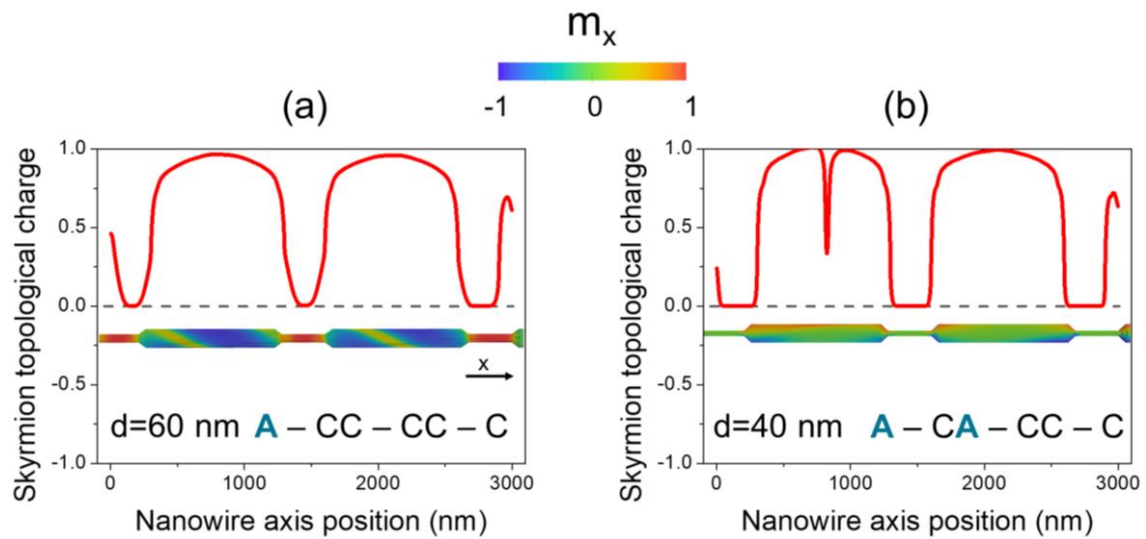

**Figure 14.** Topological charge for magnetization structure in the cork-screw regime for multisegmented FeCo nanowires, evaluated along the nanowire. The letter C stands for the clock-wise chirality of the vortex/skyrmion formed at the constriction or end of the nanowire while the letter A denotes an anti-clockwise chirality. The colored insert show the surface magnetization distribution with a chiral structure.

Once the demagnetization is completed in the large diameter segments, the skyrmion structure remains pinned at the constrictions. Importantly, the skyrmion center along the segment is not positioned in the center of nanowires but forms a spiral, see Fig. 13a and inserts in Figs 14 called in Ref.[Fernandez-Roldan2018a] a core-screw. The occurrence of spiral for vortex/skyrmion tube center is a consequence of the minimization of magnetostatic charges and is a precursor of the magnetic domains. These charges are typically created at the constrictions and are re-distributed along the nanowire length to minimize the energy. Ref.[Arrott2016] argues that these charges are proper to all magnetic nanowires. But they should be especially visible in nanowires with large saturation magnetization where the minimization of magnetostatic energy may create structures following this pattern. In our particular case the presence of the constrictions increases the effect since the vortex/tube core is additionally displaced from the center at the constriction to minimize the energy. In simulations the spiral ("corkscrew") may also appear in straight magnetic nanowires with diameters ca. 100 nm and larger, especially for nanowires with large magnetization saturation value. Similar topologically protected structures and the core-screw surface structure have been recently reported in FeNi nanowires with chemical barriers [Ruiz-Gomez2018]. Recently, they were also reported in simulations of magnetization distribution and observed by Kerr microscopy in microwires Ref. [Chizhik2018]. However, experimental verification of the "core screw" is needed and requires 3D imaging techniques.

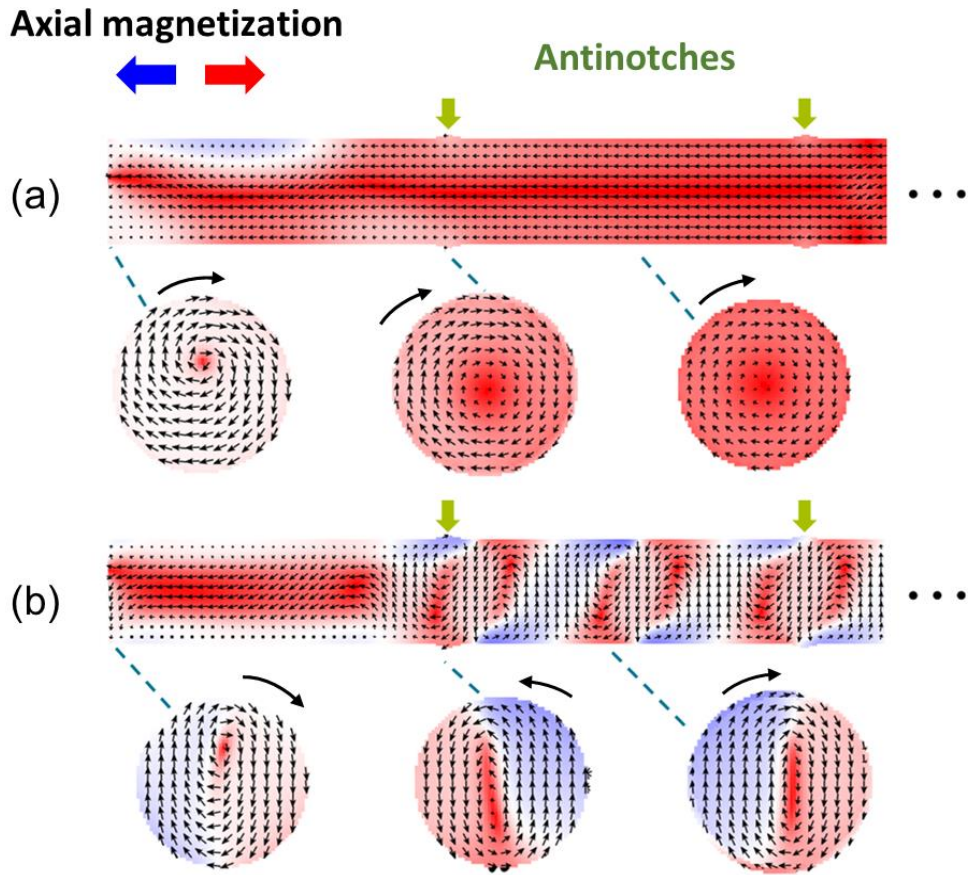

**Figure 15.** Simulated magnetic configurations of (a) FeCo and (b) Co bamboo nanowires at remanence.

The figures show a half section from the total length of the wires, containing an edge and two modulations marked by green arrows. Below each nanowire, transverse cross sections at the marked regions along the length are presented. The black arrows denote the projection of the magnetization in the corresponding section while the red-white-blue color gradient of the background corresponds to the longitudinal magnetization. Figures readapted from Ref. [Bran2016]. Modelling parameters are collected in Table I.

Fig.15 presents simulated magnetization configurations in a "bamboo" (with antinotches) nanowire with FeCo (a) and hcp Co (b) compositions in the remanent state. In the case of FeCo nanowire we observe an overall axial orientation of the magnetic moments (with the exception of the regions close to the antinotches) that is principally determined by the shape anisotropy being larger than the magnetocrystalline anisotropy of FeCo. The magnetic structure corresponds to a nearly uniform axial domain state with open vortices at each nanowire end. At the positions of an antinotch (pointed by green arrows), we observe a smooth curling of magnetization at the surface of the nanowire which occurs in order to minimize the formation of magnetic poles at the modulation. This specific behavior where the magnetization is deviated from its axial orientation is in agreement with the periodic contrasts observed in MFM and XMCD measurements [Berganza2016, Bran2016] for the FeCo-based "bamboo" nanowires.

On the other hand, the magnetic configuration of the individual Co bamboo nanowire at remanence in Fig. 15(b) displays a more complex structure. Vortex structures with alternating chiralities are observed along the nanowire length as in the case of straight Co hcp nanowires, see Fig.6. The vortices show an elongated core as a result of the distortion of its shape along the easy axis of the high transverse magnetocrystalline anisotropy and displacement in the direction normal to the easy axis. In addition, the vortex cores are not at the nanowire center but again form a spiral structure, similar to Fig.13a. Finally, micromagnetic calculations inform about the lack of correlation between the multi-vortex structure and the geometric modulations. These nanowires were studied by MFM and XMCD in Ref. [Bran2016] and micromagnetic simulations here provided complimentary information helping to interpret the images.

**5. Domain walls and domains in multisegmented Co/Ni nanowires.**

Another approach to create the energy landscape needed to control the DW along the length of the cylindrical nanowire was reported in the Refs. [Ivanov2016, Berganza2017] with alternative materials aimed at creating periodically modulated anisotropy. Co and Ni were chosen due to distinct difference in the magnetization saturation and magneto-crystalline anisotropy [Ivanov2013]. Additionally, it has been shown the nanowires of only Co or Ni are grown perfectly by the electrodeposition in alumina templates with large grain size for Ni and being monocrystalline for Co [Vilanova2015, Ivanov2013b, Ivanov2016b]

Short-segmented nanowires (with lengths ca. 100 nm) may be used to realize the concept of the "bar code" see Fig.16.

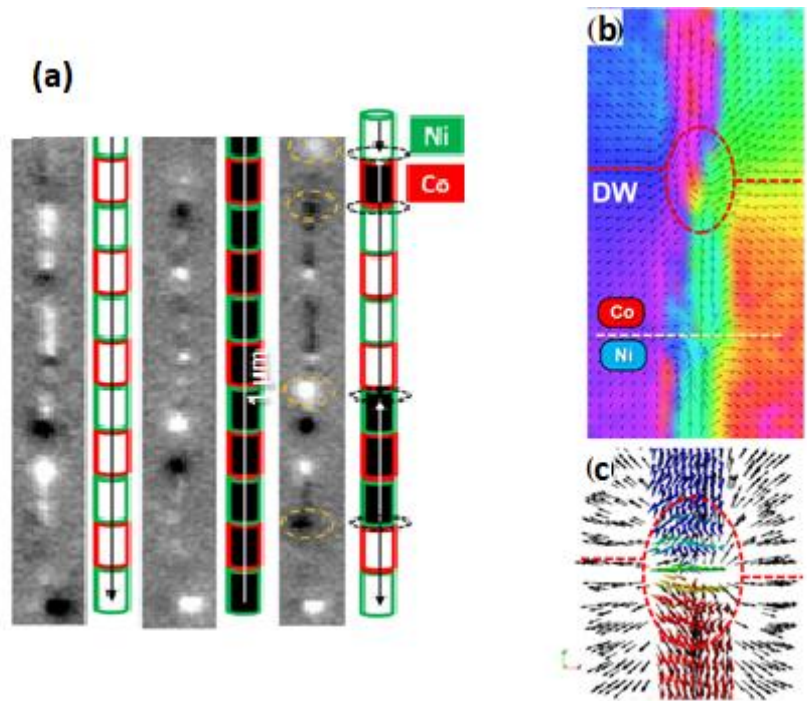

**Figure 16.** (a) Magnetic Force microscopy images of a multisegmented 80 nm diameter Co/Ni NW showing pinned magnetic structures (b) Magnified virtual bright field differential phase contrast transmission electron microsopyc image of the vortex domain wall pinned at the Co/Ni interface. The arrows show the direction of the magnetic field, which is constructed from the raw data of two orthogonal components of the magnetic field. The dashed lines show the position of the Co/Ni interface (purple) and domain wall (red). (c) Micromagnetic simulation of the pinned domain wall structure at the Co/Ni interface (color cones show magnetization direction; black arrows are calculated stray fields). Modelling parameters are collected in Table I.

Micromagnetic simulation (Fig.16c) of the nanowire consisting of two segments of Co and Ni show that the Co/Ni interface is able effectively pin the domain wall during magnetization reversal process [Mohammed2016]. The simulation proceeds from a saturated state of the multisegmented NW. Upon decreasing the field, open vortex states form at the ends of the segments at the remanence. During further increasing of the external field in the opposite direction of the initial magnetization, a subsequent reversal of magnetization of the Ni segment was observed. At this point of reversal of the multi segmented NWs, we note that the Ni segment has switched its magnetization whereas the Co segment remains relatively unchanged. This differential switching of the segments lead to the pinning of a domain wall at the Co/Ni interface. The MFM and TEM experiments shown on the Figure 16 (a,b) confirmed the pinning efficiency of the Co/Ni interfaces. The multisegmented Co/Ni nanowires demonstrated periodic pattern of the strong stray fields emanating from interfaces between Co and Ni. Such field acts as a pinning center for the DW.

Another key point is the nature of the domain wall studied in the multisegmented Co/Ni nanowires. The structure of the domain walls inside 80 nm Co or Ni nanowire should correspond to Bloch point vortex domain wall [Ivanov2013]. Figure 16(c) shows the 3D spin configuration of the domain wall in Co/Ni nanowire extracted from the micromagnetic simulations together with the calculated stray fields. The very good comparison with the experimentally observed magnetic induction map (Figure 16b) is clear.

A different possibility is offered by longer segments of Co and Ni nanowires.
In Ref. [Berganza2017] CoNi/Ni multisegmented nanowires of ca. 1μm segment lengths were studies by means of the MFM and interpreted by means of micromagnetic modeling. Importantly, $Co_{65}Ni_{35}$ segments were grown in the fcc structure on fcc Ni. However, the difference in the saturation magnetization promotes vortex structure in CoNi segments wile Ni is always in the axial domain state, see Fig. 17 below. Importantly, the transition between single vortex and multivortex state in CoNi can be controlled by fields perpendicular to nanowire axis.

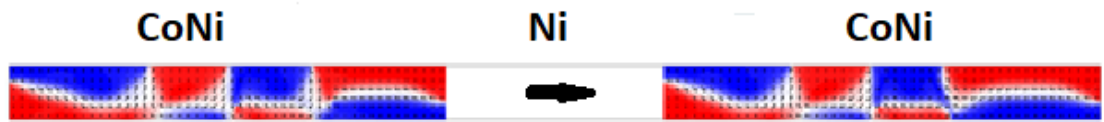

**Figure 17.** One of the possible magnetization configurations simulated in $Co_{65}Ni_{35}$/Ni multisegmented nanowires showing axial magnetization configurations for Ni and multivortex state for CoNi segments. Two segments of 1μm long CoNi separated by a Ni segment of 500 nm were considered. The diameter is 120nm and we have considered the fcc texture both in Ni and CoNi, grown in the <110> direction.

**Conclusions**

Micromagnetic simulations offer important possibilities to design magnetic cylindrical nanowire with the aim to control the type of magnetic domains, domain wall as well as their pinning mechanisms which may serve as a basis for future nanotechnological applications of these objects. They also offer a complementary tool for interpretation of magnetic imaging techniques and are typically in agreement with them provided the crystallographic structure is known in sufficient details. Domain walls type in nanowires are very rich and probably not all of them are reported and studies yet. Their occurrence is magnetic and geometrical parameters dependent. Cylindrical geometry produces a plethora of non-trivial topologies for magnetic structures which will be the subject of future studies. The change of magnetic parameters and geometries of additional possibilities to control domain wall structures and their pinning. The topological effects should play an important role in the magnetic response of nanowires. Here we have presented an example of the magnetization pinning in high-saturation FeCo multisegmented nanowire where skyrmion tubes with a curling core structure are formed. Other non-trivial structures will probably be discovered in the future. Importantly, domain walls in cylindrical nanowires are expected to have very high velocities, comparable to those expected from antiferromagnetic spintronics [Gomonay2016]. Magnetic structures in cylindrical nanowires also may have interesting dynamical properties coming from their topology and emerging electrodynamics effect may appear [Charilaou2018] We anticipate that some of these effects may found important nanotechnological applications in the future.